\documentstyle[11pt,epsfig]{article}
\begin{document}
\baselineskip=23pt
 ~~~~~~~~~~~~~~~~~~~~~~~~~~~~~~~~~~~~~~~~~~~~~~~~~~~~~~~~~~~~~~~~~~~~~~~~~~~~~~~~~~~~~~~~~~~~BIHEP-TH-2002-53

\vspace{1.2cm}

\begin{center}
{\Large \bf Noncommutativity as a Possible Origin of the Ultrahigh
Energy Cosmic Ray and the TeV-photon Paradoxes }

\bigskip

Shao-Xia Chen\footnote{ruxanna@mail.ihep.ac.cn}~~and~~Zhao-Yu Yang\footnote{yangzhy@mail.ihep.ac.cn}\\
{\em Institute of High Energy Physics,
Chinese Academy of Sciences} \\
{\em P.O.Box 918(4), 100039 Beijing, China}
\end{center}

\vspace{1.2cm}
\begin{abstract}
In this paper, we present a general modified dispersion relation
derived from $q$-deformed noncommutative theory and apply it to
the ultrahigh energy cosmic ray and the TeV-photon
paradoxes---threshold anomalies. Our purpose is not only trying to
solve these puzzles by noncommutative theory but also to support
noncommutative theory through the coincidence of the region in the
parameter space for resolving the threshold anomalies with the one
from the $q$-deformed noncommutative theory.
\end{abstract}

\vspace{1.2cm}
\vspace{1.2cm}
\section{Introduction}
Recently there is great interest in the study of the ultrahigh
energy cosmic ray (UHECR) and the TeV-photon paradoxes. The first
paradox is that some experiments \cite{data1}-\cite{data7}
observed many hundreds of events with energies above $10^{19}$eV
and about 20 events above $10^{20}$eV which are above the
Greisen-Zatsepin-Kuzmin (GZK) threshold. In principle, photopion
production with the microwave background radiation photons should
decrease the energies of these protons to the level below the
corresponding threshold. The second paradox is the fact that
experiments detected 20TeV photons from Mrk 501 (a BL Lac object
at a distance of 150Mpc). Similarly to the first case, due to the
interaction with the IR background photons, the 20TeV photons
should have disappeared in the ground-based detections. These two
puzzles have a common feature that both of them can be seen as
some threshold anomalies: energy of an expected threshold is
reached but the threshold is not observed. There are numerous
solutions such as \cite{UHECR, solution1} proposed for the UHECR
and the TeV-$\gamma$ paradoxes. In particular, most authors
\cite{loren1}-\cite{loren8} have suggested that Lorentz-invariance
violation can be the origin of these anomalies and have obtained
many developments.

In this paper, we try to deal with these paradoxes by a modified
dispersion relation which is obtained from $q$-deformed
noncommutative theory. It is well-known that noncommutative
geometry \cite{non-cones, non} plays an important role in the
trans-Planck physics, therefore, for phenomena related to the
ultrahigh energy physics such as these two paradoxes, it is more
reasonable to consider the Lorentz-invariance violation is from
noncommutative geometry and make some explanations for them in the
background of noncommutative theory. The possibility has been
mentioned that noncommutative geometry may be responsible for the
ultrahigh energy cosmic ray and TeV-photon paradoxes in some
papers\cite{a2}. Here through a complete process based on a
brand-new method which is from $q$-deformed noncommutative theory,
we will connect threshold anomalies with noncommutative theory to
not only unravel these threshold anomalies but also present a
strong support for noncommutative theory in turn.

Our paper is organized as follows: We derive a general modified
dispersion relation from $q$-deformed noncommutative theory in
section 2. In the following section we will apply this dispersion
relation to the ultrahigh energy cosmic ray and the TeV-photon
paradoxes, and with the observed energies we get a common region
of the ($\omega$,~$|q-1|$) parameter space to resolve these two
anomalies. In section 4, we will give some implications and
remarks for our work.

\section{A General Modified Dispersion Relation from $q$-deformed Noncommutative Theory}
We start our discussion from the general dispersion relation for a
massive particle ($m=0$ for a massless particle) in the realistic
case (from then on, we will use the notation $\hbar=c=1$),
\begin{equation}
\label{dispersion} E^2=m^2+p^2.
\end{equation}
The Hamiltonian operator for a fermion in the ordinary case is:
\begin{equation}
\label{classical}
\hat{H}=\frac{1}{2}\omega\left(\hat{a}^\dag\hat{a}-\hat{a}\hat{a}^\dagger\right).
\end{equation}
The number operator is defined as:
\begin{equation}
\label{1}
\hat{a}^\dag\hat{a}=\hat{N},~~~\hat{a}\hat{a}^\dagger=\widehat{1-N}.
\end{equation}
So we have:
\begin{equation}
\label{2}
\hat{H}=\frac{1}{2}\omega\left(\hat{N}-(\widehat{1-N})\right).
\end{equation}
Then
\begin{equation}
\label{epsilon-n}
E=\frac{1}{2}\omega\left(n-(1-n)\right)=\frac{1}{2}\omega\left(2n-1\right).
\end{equation}
From (\ref{dispersion}) and (\ref{epsilon-n}), we can easily
derive:
\begin{equation}
\label{fer-n-p} n=\frac{\sqrt{m^2+p^2}}{\omega}+\frac{1}{2}.
\end{equation}
And in the $q$-deformed case \cite{ng} defined,
\begin{equation} \label{general}
\hat{a}^\dag_q\hat{a}_q=[\hat{N}]_q,~~~\hat{a}_q\hat{a}_q^\dagger=[\widehat{1-N}]_q,~~~[x]_q\equiv\frac{q^x-q^{-x}}{q-q^{-1}},
\end{equation}
where $q$ is a complex deformation parameter. The non-deformed
case is obtained by setting $q$ equal to 1.

Therefore, from \cite{su}, we can directly get:
\begin{equation}
\label{fermi-H-n}
\hat{H}_q=\frac{\omega}{2}\left([\hat{N}]_q-[\widehat{1-N}]_q\right).
\end{equation}
Applying (\ref{fer-n-p}) to (\ref{fermi-H-n}), we can obtain
\begin{equation}
\label{q-fermi-dispersion}
E=\frac{\omega}{2}\left(\left[\frac{\sqrt{m^2+p^2}}{\omega}+\frac{1}{2}\right]_q-\left[\frac{1}{2}-\frac{\sqrt{m^2+p^2}}{\omega}\right]_q\right).
\end{equation}

For a boson in the non-deformed case:
\begin{equation}
\label{bose-h-n}
\hat{H}=\frac{1}{2}\omega\left(\hat{b}^\dag\hat{b}+\hat{b}\hat{b}^\dagger\right),
\end{equation}
\begin{equation}
\label{bose-n-b}
\hat{b}^\dag\hat{b}=\hat{N},~~~\hat{b}\hat{b}^\dagger=\widehat{N+1}.
\end{equation}
The $q$-deformed number operator was defined in
\cite{bose-number1-q-H-n, bose-number2, bose-number3}:
\begin{equation}
\label{bose-n-b-deform}
\hat{b}^\dag_q\hat{b}_q=[\hat{N}]_q,~~~\hat{b}_q\hat{b}^\dagger_q=[\widehat{N+1}]_q,
\end{equation}
and the $q$-deformed Hamiltonian \cite{bose-number1-q-H-n,
0205208} is:
\begin{equation}
\label{qH}
\hat{H}_q=\frac{\omega}{2}\left([\hat{N}]_q+[\hat{N}+1]_q\right).
\end{equation}
Similarly to the case of a fermion, we obtain the $q$-deformed
dispersion relation of a boson:
\begin{equation}
\label{q-bose-dispersion}
E=\frac{\omega}{2}\left(\left[\frac{\sqrt{m^2+p^2}}{\omega}-\frac{1}{2}\right]_q+\left[\frac{\sqrt{m^2+p^2}}{\omega}+\frac{1}{2}\right]_q\right).
\end{equation}
From (\ref{q-fermi-dispersion}), (\ref{q-bose-dispersion}) and
according to the definition of $[x]_q$, we have the same
dispersion relation for a boson and a fermion in the $q$-deformed
case.

 Expand the right hand side of
(\ref{q-fermi-dispersion}) and (\ref{q-bose-dispersion}) in the
neighborhood of $q=1$ to the second order, we derive the common
approximate deformed dispersion relation for a fermion and a
boson:
\begin{equation}
\label{q-dispersion}
E=\sqrt{m^2+p^2}+\frac{\sqrt{m^2+p^2}(4m^2+4p^2-\omega^2)}{24\omega^2}(q-1)^2.
\end{equation}

\section{Discussion for Threshold Anomalies in the UHECR and the TeV-photon Paradoxes}
Now we will compute the $q$-deformed threshold for the UHECRs and
the TeV-photons. We consider the head-on collision between a soft
photon of energy $\epsilon$, momentum $\vec{k}$ and a high energy
particle of energy $E_1$, momentum $\vec{p}_1$, which leads to the
production of two particles with energies $E_2$, $E_3$ and momenta
$\vec{p}_2$, $\vec{p}_3$ respectively \cite{UHECR}. From the
energy conservation and momentum conservation, we have:
\begin{equation}
\label{threshold1} E_1+\epsilon=E_2+E_3;
\end{equation}
\begin{equation}
\label{threshold2} p_1-k=p_2+p_3.
\end{equation}
In the C. O. M. frame, $m_2$ and $m_3$ are rest at threshold, so
they have the same velocity in the lab frame. It's easy to give
the following equation:
\begin{equation}
\label{p3p2} \frac{p_2}{p_3}=\frac{m_2}{m_3}.
\end{equation}
Applying our resulting dispersion relation (\ref{q-dispersion})
and (\ref{p3p2}) to (\ref{threshold1}) and (\ref{threshold2}), we
can obtain the expression for $p_{1,th,{\rm q}}$ in this case:
\begin{equation}
\label{q-threshold} p_{1,th,{\rm q
}}=\frac{(m_2+m_3)^2-m_1^2}{4\epsilon}+\frac{p_{1,th,{\rm q
}}^4}{12\epsilon\omega^2}\left(\frac{m_2^3+m^3_3}{(m_2+m_3)^3}-1\right)(q-1)^2.
\end{equation}
In the course of deriving (\ref{q-threshold}), we have used the
simplified form of (\ref{q-dispersion}):
\begin{equation}
\label{simplified-dispersion}
E_i=p_i+\frac{m_i^2}{2p_i}+\left(p_i+\frac{m_i^2}{2p_i}\right)\left(\frac{1}{6\omega^2}\left(p_i^2+m_i^2\right)-\frac{1}{24}\right)(q-1)^2,
\end{equation}
\begin{equation}
\epsilon=k+k\left(\frac{k^2}{6\omega^2}-\frac{1}{24}\right)(q-1)^2.
\end{equation}
While in the non-deformed case, the threshold has the form as
below:
\begin{equation}
\label{nq-threshold} p_{1,th,{\rm
nq}}=\frac{(m_2+m_3)^2-m_1^2}{4\epsilon}.
\end{equation}
Hereafter, we will denote the $p_{1,th}$ in the non-deformed case
and in the $q$-deformed case by $p_{1,th,{\rm nq}}$ and
$p_{1,th,{\rm q}}$ separately.

In the UHECR paradox $p+\gamma\rightarrow p+\pi$, we set
$m_1=m_2=m_p=940{\rm MeV},~m_3=m_{\pi}=140{\rm MeV}$, in the other
one $\gamma+\gamma\rightarrow e^++e^-$, we have $m_1=0$ and
similarly set $m_2=m_3=m_e=0.5{\rm MeV}$. Inputting the values of
$p_{1,th}$ in \cite{UHECR}
\begin{equation}
\label{UH-threshold} p_{1,th,{\rm nq}}^{{\rm UHECR}}=5\times
10^{19}{\rm eV},~~p_{1,th,{\rm q}}^{{\rm UHECR}}=3\times
10^{20}{\rm eV},
\end{equation}
and
\begin{equation}
\label{gamma-threshold} p_{1,th,{\rm nq}}^{\gamma}=10{\rm
TeV}~~,~~ p_{1,th,{\rm q}}^{\gamma}=20{\rm TeV}
\end{equation}
to (\ref{q-threshold}) and (\ref{nq-threshold}) respectively, we
can get two curves for the relation between $\omega$ and $|q-1|$
as plotted in Fig. 1 (the unit of $\omega$ is MeV).
\begin{center}
\begin{figure}[h]
\psfig{file=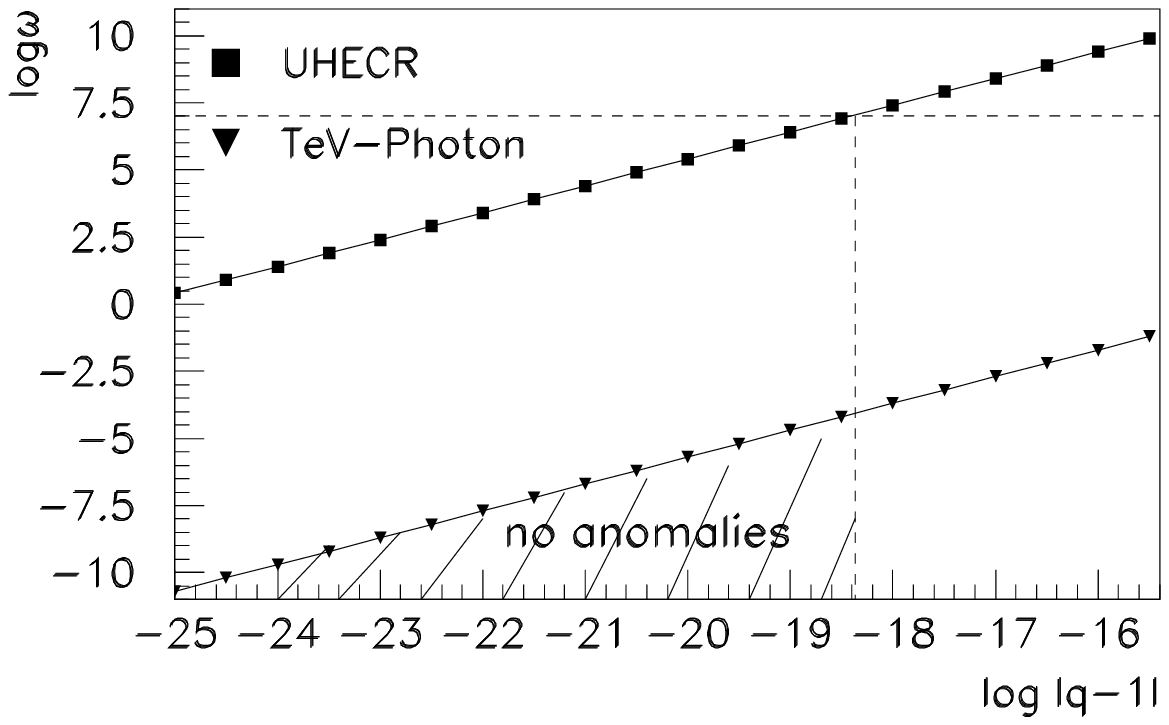,bbllx=2.cm,bblly=12.cm,bburx=14.cm,bbury=18.cm,height=6.cm,width=12.cm,angle=0}
\end{figure}
\end{center}
\centerline{Figure 1: \bf the region of parameter space}

The region of the $(\omega,~|q-1|)$ parameter space provides a
solution to both the UHECR and the TeV-$\gamma$ threshold
anomalies. While applying the $q$-deformed dispersion relation to
the present observed anomalies, the relation between $\omega$ and
$|q-1|$ can be described by the two curves in Fig. 1 which give
the lower bound on the $q$-deformation. In the region below both
of the curves, the UHECR and the TeV-photon paradoxes are resolved
together. The top left corner is excluded, which is easily
understood for at a fixed $\omega$ the conventional dispersion
relation can be reproduced if we take $|q-1|\rightarrow 0$. Under
the consideration of $\omega<10^7$ MeV, the parameters must be
within the shaded area. $\omega<10^7$ MeV is set according to the
smallness of $|q-1|$, which is the requirement of the $q$-deformed
noncommutative theory.

\section{Implications and Remarks}
The relation between the two parameters $\omega$ and $q$ depends
strongly on the experimental input. In this paper we just made use
of the highest energies of the ultrahigh energy cosmic rays and
the TeV-photons arriving on earth observed so far. Because both
threshold anomalies of the UHECRs and the Mrk 501 TeV-photons are
still under the situation that the data should be proceeded very
cautiously, our resulting relation of the parameters is to be
tested. If in the future better data prove the assumption of these
threshold anomalies is incorrect, the explanation for them in
terms of the $q$-deformed noncommutative theory has to be
excluded, the same case as other interpretations such as the
quantum-gravity-motivated models \cite{UHECR, uncertainty} will
encounter. If, however, the threshold anomalies are confirmed, the
enhanced constraints on the available region of the parameter
space can result from future cosmic-ray observations and
experiments performed at very high engines.

On the other hand, it has long been recognized that cosmology is a
natural laboratory for testing the possible theories of extremely
high energy physics such as string theory and noncommutative
theory, and many authors \cite{0205208, string-cos1, string-cos2,
non-cos1, non-cos2, non-cos3} have tried to search for some
supports for these theories from the cosmological observations and
experiments directly or indirectly. One purpose of this paper is
also following this train of thought, and if it can be verified
that the permitted region of the parameter space to resolve the
threshold anomalies is coincident with the parameter range from
the noncommutative theory, it will be more confident for us to
consider the noncommutative theory as the suitable theory for the
trans-Planck physics. \vspace{0.5cm}

\centerline{\large\bf Acknowledgements} We thank very much to Dr.
G. Amelino-Camelia for valuable suggestions and comments. The work
was supported in part by the Natural Science Foundation of China.


\begin{thebibliography}{999}
\bibitem{data1} M. Takeda {\it et al}., Phys. Rev. Lett. {\bf 81}, 1163 (1998).
\bibitem{data2} M. Takeda {\it et al}., arXiv: astro-ph/9902239.
\bibitem{data3} N. Hayashida {\it et al}., Phys. Rev. Lett. {\bf 73}, 3491 (1994).
\bibitem{data4} D. J. Bird {\it et al}., Astrophys. J. {\bf 441}, 144 (1995); D. J. Bird {\it et al}., Phys. Rev. Lett. {\bf 71}, 3401 (1993); D. J. Bird {\it et al}., Astrophys. J. {\bf 424}, 491 (1994).
\bibitem{data5} M. A. Lawrence, R. J. O. Reid and A. A. Watson, J. Phys. G {\bf 17}, 773 (1991).
\bibitem{data6} N. N. Efimov {\it et al}., Ref. Proc. International
Symposium on {\it Astrophysical Aspects of the Most Energetic
Cosmic Rays}, World Scientific, Singapore, pp. 20, (1991).
\bibitem{data7} D. Kieda {\it et al}., HiRes Collaboration,
Proceeds. 26th ICRC (Salt Lake 1999).
\bibitem{UHECR} G. Amelino-Camelia and T. Piran, Phys. Rev. D {\bf 64}, 036005 (2001).
\bibitem{solution1} J. M. Carmona, J. L. Cort$\acute{{\rm e}}$s, J. Gamboa
and F. M$\acute{{\rm e}}$ndez, arXiv: hep-th/0207158.
\bibitem{loren1} L. G. Mestres, arXiv: physics/9704017.
\bibitem{loren2} S. Coleman and S. L. Glashow, Phys. Rev. D {\bf
59}, 116008 (1999).
\bibitem{loren3} R. Aloisio, P. Blasi, P. L. Ghia, and A. F. Grillo, Phys.
Rev. D {\bf 62}, 053010 (2000).
\bibitem{loren4} O. Bertolami and C. S. Carvalho, Phys. Rev. D
{\bf 61}, 103002 (2000).
\bibitem{loren5} H. Sato, arXiv: astro-ph/0005218.
\bibitem{loren6} T. Kifune, Astrophys. J. {\bf 518}, L21
(1999).
\bibitem{loren7} W. Kluzniak, arXiv: astro-ph/9905308.
\bibitem{loren8} R. J. Protheroe and H. Meyer, Phys. Lett. B {\bf
493}, 1 (2000).
\bibitem{non-cones} A. Connes, {\em Noncommutative Geometry}, Academic Press, London (1994).
\bibitem{non} J. Wess, Int. J. Mod. Phys. A {\bf 12}, 4997 (1997).
\bibitem{a2} G. Amelino-Camelia, Int. J. Mod. Phys. D {\bf 11}, 35
(2002).
\bibitem{ng} Y. J. Ng, J. Phys. A {\bf 23}, 1023 (1990).
\bibitem{su} M. L. Ge and G. Su, J. Phys. A {\bf 24},
L721 (1991).
\bibitem{bose-number1-q-H-n} M. A. M. Delgado, J. Phys. A {\bf
24},
L1285 (1991).
\bibitem{bose-number2} S. Vokos and C. Zachos, Mod. Phys. Lett. A
{\bf 9}, 1 (1994).
\bibitem{bose-number3} J. A. Tuszy$\acute{{\rm n}}$ski, J. L.
Rubin, J. Meyer, and M. Kibler, Phys. Lett. A {\bf 175}, 173
(1993).
\bibitem{0205208} Z. Chang and S. X. Chen, arXiv: cond-mat/0205208.
\bibitem{uncertainty} Y. J. Ng, D. S. Lee, M. C. Oh, and H. van Dam,
Phys. Lett. B {\bf 507}, 236 (2001).
\bibitem{string-cos1} J. A. Gu, P. M. Ho and S. Ramgoolam, arXiv:
hep-th/0101058.
\bibitem{string-cos2} S. F. Hassan and M. S. Sloth, arXiv:
hep-th/0204110.
\bibitem{non-cos1} C. S. Chu, B. R. Greene and G. Shiu, Mod. Phys.
Lett. A {\bf 16}, 2231 (2001).
\bibitem{non-cos2} F. Lizzi, G. Mangano, G. Miele, and M. Peloso, JHEP {\bf 0206}, 049
(2002).
\bibitem{non-cos3} R. H. Brandenberger and P. M. Ho, Phys. Rev. D {\bf 66}, 023517
(2002).


\end{thebibliography}
\end{document}